\documentclass[12pt,colordvi,newlfont,times,rotating,epsfig]{article}

\usepackage{wrapfig,epsfig}

\newcommand{\bfg}{\begin{figure}}
\newcommand{\efg}{\end{figure}}
\newlength{\dinwidth}                       
\newlength{\dinmargin}                      
\def\lsim{\mathrel{\rlap{\lower4pt\hbox{\hskip1pt$\sim$}}
    \raise1pt\hbox{$<$}}}                
\def\gsim{\mathrel{\rlap{\lower4pt\hbox{\hskip1pt$\sim$}}
    \raise1pt\hbox{$>$}}}                
\def\3{\ss}

\def\beq{\begin{equation}}
\def\eeq{\end{equation}}
\begin{document}
{\flushleft DESY 98-060\\May 1998} 
\vspace*{40mm}
\begin{center}
\large{ \bf{RECENT HERA RESULTS AND FUTURE PROSPECTS \\}}
\end{center}

\vspace*{5mm}

\begin{center}
Uwe Schneekloth \\
Deutsches Elektronen Synchrotron, Notkestrasse 85, 22603 Hamburg 
\end{center}
%
\vspace*{60mm}
\begin{center} \bf{ Abstract} \end{center}
\noindent
A few selected HERA results are presented and the prospects for 
 future measurements with high luminosity  are discussed,  
which will become available after the planned luminosity upgrade 
of the HERA storage ring planned for 2000. 

\newpage
\section{Present HERA Performance}

\begin{wrapfigure}{r}{8cm}
\centerline{\epsfig{figure=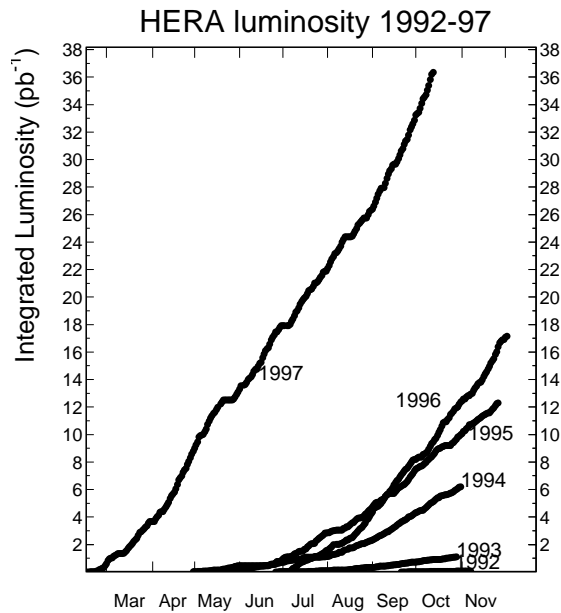,%
             bbllx=65,bblly=170,bburx=520,bbury=680,width=8cm}}
\caption{\sf The integrated luminosity delivered by HERA 
to the ZEUS interaction region for the different years 
of operation versus the date in that year. }
\label{lumi}
\end{wrapfigure}
The electron - proton storage ring HERA at DESY, colliding 27.5\,GeV 
electrons or positrons on 820\,GeV protons, 
started operation in 1992. 
The instantaneous and integrated luminosities have been steadily increased 
as shown in Fig.~\ref{lumi}. 
During the first two years of operation, when electrons were colliding on 
protons,  
the electron beam current was limited by breakdowns of the beam life time. 
These limitations were not present after the machine switched to  positrons 
during the 1993/1994 winter shut-down. 
Operation in 1997 was very successful providing an increased instantaneous 
luminosity and a long duration of the running period. 
In total  integrated luminosities of 2\,pb$^{-1}$ (e$^-$ p collisions, 
1992 and 1993) and 71\,pb$^{-1}$ (e$^+$ p collisions, since 1994) 
have been delivered to the colliding beam experiments H1 and ZEUS. 

Table~\ref{para} compares the machine parameters which were achieved in 1997 
 to the original design parameters. 
The maximum luminosity  of  
$1.4\times10^{31}$\,cm$^{-2}$  s$^{-1}$ is close to the 
original design goal ($1.5\times10^{31}$\,cm$^{-2}$  s$^{-1}$).
%

The accelerator is now operated in a two year cycle. 
Short winter shut-downs of one to two months in 1996/97 and 1998/99 
alternate with long winter shut-downs in the following year. 
The major activity for HERA during the present winter shut-down is a 
modification of 
the vacuum system of the electron ring in order to allow 
high current electron running. 
The expected integrated luminosities per experiment ($e^- p $ collisions) 
for the next two years are about 
15\,pb$^{-1}$ in 1998 (short running period) and $>$35\,pb$^{-1}$ 
in 1999 (long running period). 
The luminosity upgrade and installation of spin rotators in the south (ZEUS)
and north (H1) straight sections are planned for the 1999/2000 winter 
shut-down.

\begin{table}[htb]
\centering
\begin{tabular}{|l||c|c|c| }
\hline
\multicolumn{4}{ |c| } {HERA  Parameters }   \\
                        & 1997    & Design & Upgrade \\ 
\hline\hline
 p beam energy (GeV)    &  820    &  820   & 820$^\star$  \\
 e beam energy (GeV)    &  27.5   &  30    &  30$^\dagger$  \\
 Number of bunches (proton/electron) & 180/189 & 210    & 180/189  \\
 Number of protons/bunch  & $7.7\times10^{10}$ & $10\times10^{10}$ 
         &                 $10\times10^{10}$  \\
 Number of electrons/bunch & $2.9\times10^{10}$ & $3.6\times10^{10}
 $             & $4.2\times10^{10}$ \\
 Proton current (mA)      & 105   & 160    & 140      \\
 Electron current (mA)    & 43    & 58     & 58       \\
 Hor. proton emittance (nm rad)   &  5.5   &  5.7  &  5.7  \\
 Hor. electron emittance (nm rad) &  40    &  39   &  22   \\
 Proton beta function $x/y$ (m) & 7/0.5 & 10/1     & 2.45/0.18 \\
 Electron beta function $x/y$ (m) & 1/0.7 & 2/0.7  & 0.63/0.26 \\
 beam size $\sigma_x \times \sigma_y $ ($\mu$m) & 
      200 $\times$ 54   &   247  $\times$ 78  &   118  
               $\times$ 32       \\
 Synchrotron radiation at IP (kW) & 6.9&  9.7   & 25       \\
\hline
 Specific luminosity  (cm$^{-2}$  s$^{-1}$ mA$^{-2}$) &
             $7.6\times10^{29}$  & $3.4\times10^{29}$ & $1.6\times10^{30}$  \\
 Luminosity  (cm$^{-2}$  s$^{-1}$)  &
             $1.4\times10^{31}$  & $1.5\times10^{31}$ & $7.4\times10^{31}$  \\
\hline
\end{tabular}
\label{para}
\caption{\sf HERA parameters achieved in 1997, compared to the original design 
         and the luminosity upgrade program.
         $^\star$ Increase to 920\,GeV is being studied.  
         $^\dagger$ Maximum energy assumed for layout of~IP. }
\end{table}

\section{The HERA Luminosity Upgrade}


The aim of the HERA luminosity upgrade is to increase the design luminosity 
from $1.5\times10^{31}$\,cm$^{-2}$  s$^{-1}$ to 
about $7\times10^{31}$\,cm$^{-2}$  s$^{-1}$ as shown in table~\ref{para}. 
An integrated luminosity of about 1\,fb$^{-1}$ per experiment is expected to 
be delivered
to the colliding beam experiments in the running period 2000 - 2005. 
The luminosity increase will be achieved by stronger focusing of both the 
electron and proton beam. 
The final focusing magnets have to be moved closer to the interaction 
point (IP).  
This necessitates an earlier separation of both beams at the interaction 
region,
which requires new superconducting beam magnets to be placed inside the H1 and 
ZEUS detectors. 
The design of the new interaction region is  challenging, because 
of a significant increase of the synchrotron radiation emitted near the IP.
Details of the luminosity upgrade can be found in~\cite{lumiup}.
In table~\ref{para} the electron beam energy of 30\,GeV was chosen as the
 maximum energy 
for background calculations and the design of the new interaction region. 
The actual electron energy will  be slightly lower, because of constraints
on the  reliability and power  of the RF system. 
There are plans for an increase of the proton beam energy from 820 to 
920\,GeV. First tests were successfully performed during the machine 
development period in  December 1997. 
If further tests scheduled during the re-commissioning in 1998 are successful 
too,
the beam energy may already be increased for the 1998 running period. 
%
%

\section{Detector Modifications} 

Both collaborations have made several modifications of their detectors during 
the last years. H1 installed a new backward tracking detector and a new 
rear calorimeter. 
Small electromagnetic calorimeters near the beam pipe (BPC) for tagging  
 very low Q$^2$ electrons were added to both detectors. 
In addition small Si trackers were placed in front of the BPCs in order 
to improve the measurement of the scattered electron. 
H1 installed a Si micro vertex detector with backward wheels. 
During the 1997/98 shut-down ZEUS implemented a forward plug calorimeter (FPC) 
for a better measurement of the proton remnant and very forward jets. 
These modifications, 
except the ZEUS FPC and the H1 barrel Si micro vertex detector, 
aim at improving the acceptance for electrons scattered at very small 
angle, and hence  extending  the kinematic region to  very 
low $x$ and $Q^2$ physics. 

Several detector modifications have been proposed for the HERA luminosity
upgrade: H1 is planning an upgrade of the central and forward tracking and
an upgrade of the trigger; ZEUS will install a Si micro vertex 
detector with forward wheels. 
Both collaborations have to modify their luminosity monitors and leading 
proton spectrometers. 
The compensating solenoids will be removed. 
The acceptance at very small angles will be slightly reduced 
because of the final focusing magnet inside the detector volumes, 
i.e.\,detector components at very small angles (BPC and FPC) will have to be 
removed.

\section{Selected Recent Results and Prospects}

\begin{wrapfigure}[36]{r}{8.5cm}
\centerline{\epsfig{file=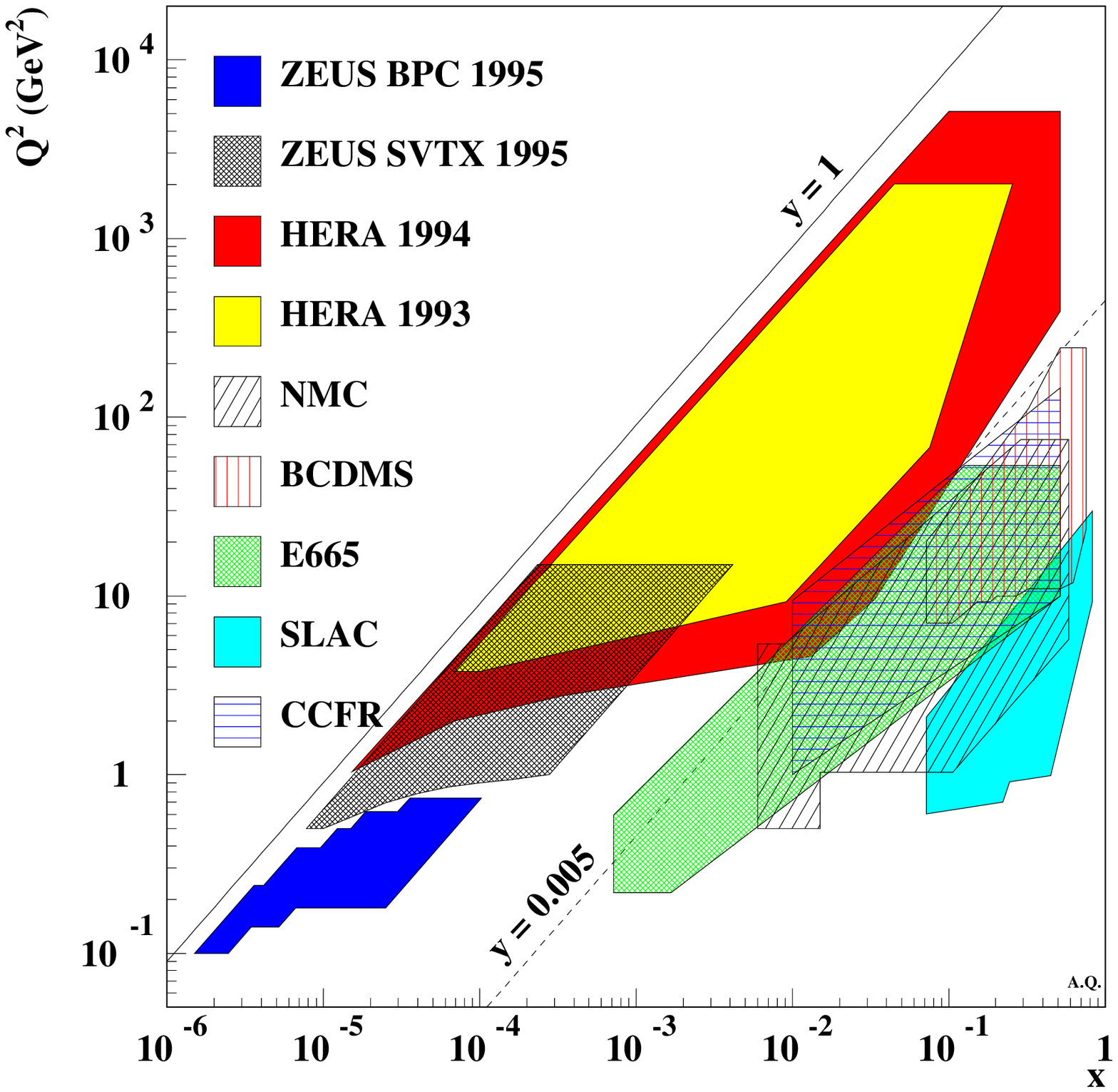,%
             bbllx=30,bblly=140,bburx=580,bbury=690,width=8.5cm,clip=}}
\caption{\sf Kinematic region of deep inelastic scattering.}
\label{kinrange}
%
\centerline{\epsfig{file=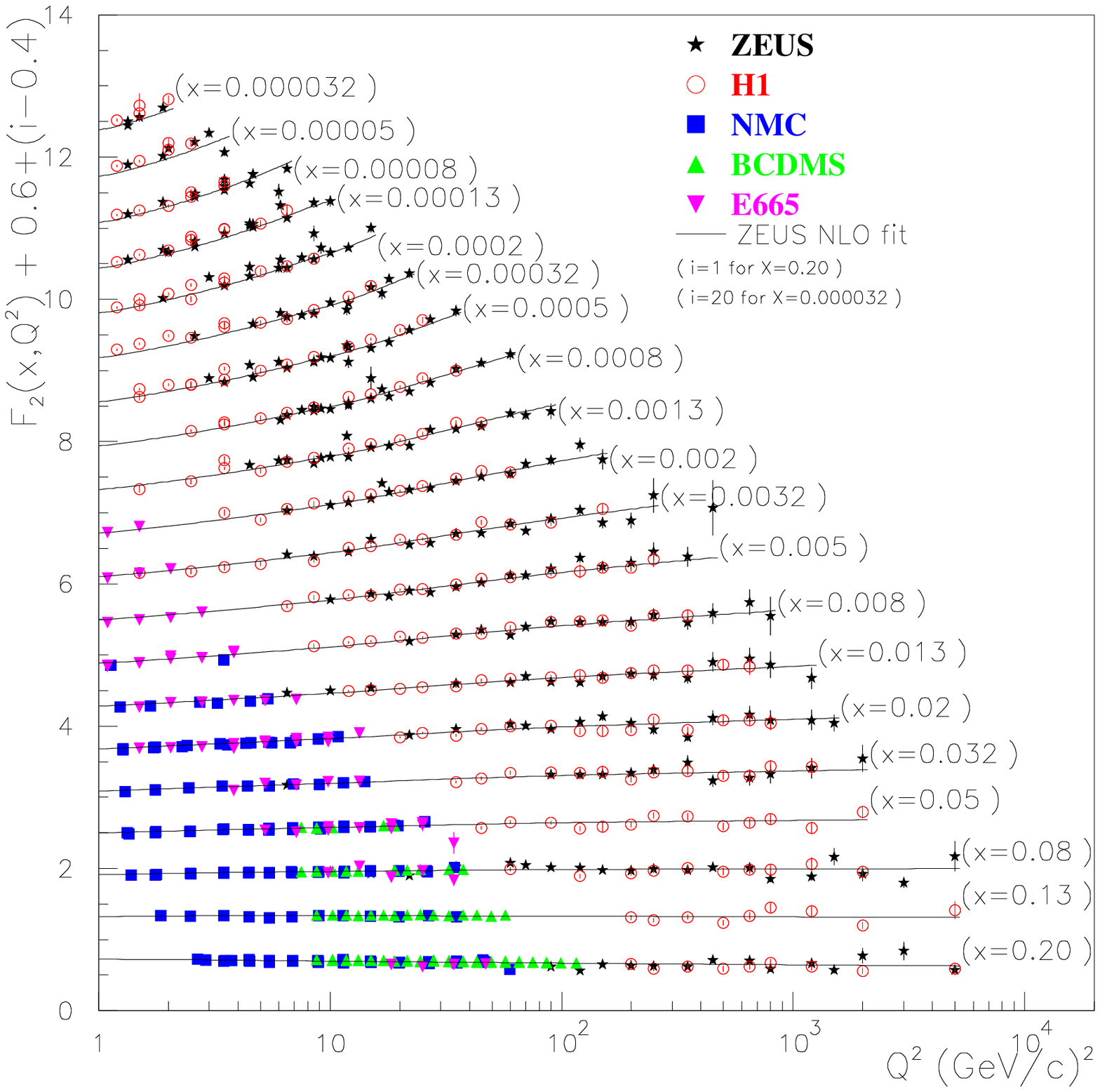,height=8cm}}
\caption{\sf Structure function $F_2$ as a function of $Q^2$ for 
         fixed values of $x$.}
\label{F2}
\end{wrapfigure}

In the following, a few selected recent results are presented and compared 
to expected future measurements, which were studied at 
the Workshop on Future Physics at HERA~\cite{workshop}. 

\subsection{Measurement of $F_2$ Structure Function}

One of the main objectives of deep inelastic scattering at HERA 
is a precision measurement of the proton structure function $F_2$ 
in a wide range  of $x$ and $Q^2$. 
The explored kinematic range in the $x,Q^2$ plane is shown in 
Fig.~\ref{kinrange} and compared to fixed target experiments. 
The HERA experiments have increased the $x$ and $Q^2$ ranges by two 
orders of magnitude each. 
During the last few years the kinematic range was extended to very low 
 $x$ and $Q^2$ by improving the detector acceptance at low scattering angles 
as described in the previous section. 
At  large $x$ and $Q^2$ the measurements are limited by the available 
statistics.  
The measured $F_2(x,Q^2)$ structure function as a function of $Q^2$ is 
shown in Fig.~\ref{F2} for  fixed values of $x$. 
Strong scaling violations are observed, which decrease as $x$ increases. 
The  data are well described by QCD using NLO DGLAP evolution in the 
full kinematic range~\cite{ZEUSF2}.

High luminosity data ($\cal {L}$ $\ge $300\,pb$^{-1}$) will provide 
a precision measurement of $F_2(x,Q^2)$ at the few percent level over 
the full accessible kinematic region from $x \sim 10^{-5}$ to 
$Q^2 \sim 50000$\,GeV$^2$. 
$F_2$ is also expected to be the  most  precise way to determine the 
strong coupling constant $\alpha_s$ and the gluon distribution $x g$ 
from $F_2$ scaling violations at the 1\% level~\cite{botje}. 

%

\subsection{Charm Contribution to Proton Structure Function}

\begin{figure}[h]
\centerline{\epsfig{file=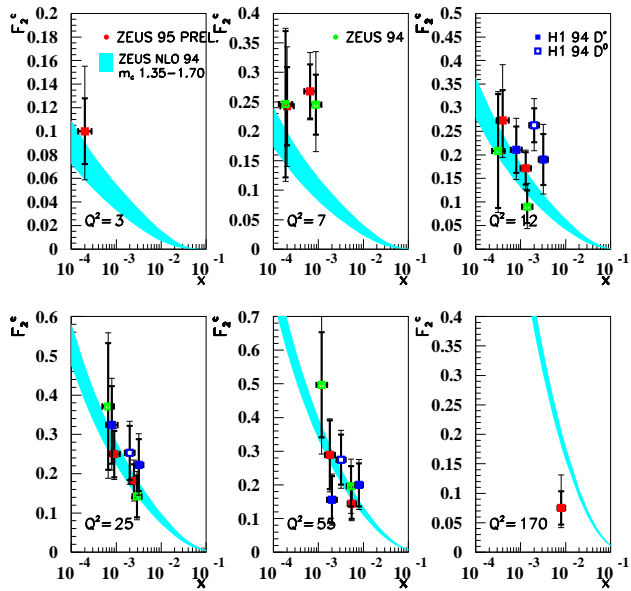,width=8cm,clip=}}
\caption{\sf $F_2^{c\bar c}$ as a function of $x$ for different $Q^2$ bins. 
         The shaded band represents the NLO calculation using the gluon 
         density extracted from the ZEUS NLO fit with  charm masses 
         ranging from 1.35 to 1.70\,GeV.}
\label{F2cc}
\end{figure}

The charm contribution to the proton structure function, $F_2^{c\bar c}$,
 can be measured directly 
by tagging charm in the hadronic final state. 
Both H1 and ZEUS so far used the decay 
$D^\star \rightarrow D^0 \pi \rightarrow K \pi \pi$ for
this measurement. 
The  differential  cross section 
$d^2 \sigma^{c\bar c} / dx dQ^2$ was measured in the 
phase space region of $| \eta^{D^\star} | < 1.5, p_t^{D^\star} > 1.5$\,GeV, 
$80 < W < 210$\,GeV and $3 < Q^2 < 170$\,GeV$^2$ (in case of ZEUS). 
The measured data are in agreement with the expectation 
of photon-gluon fusion as the primary production mechanism and 
disagree with processes in which the charmed hadrons would originate 
from charm quarks inside the proton. 
The contribution from charm to the proton
 structure function,  $F_2^{c\bar c}$, was determined from the measured 
differential  cross section 
by integrating and extrapolating outside the measured $\eta^{D^\star}$
and  $p_t^{D^\star}$ ranges. 
The results  shown in Fig.~\ref{F2cc} 
are in good agreement with  the NLO perturbative QCD calculation~\cite{NLO} 
using the gluon distributions obtained from the $F_2$ scaling 
violation of the ZEUS data~\cite{ZEUSF2} shown as band in Fig.~\ref{F2cc}.  
The $F_2^{c\bar c}$ contribution accounts for  about 25\,\% of the 
$F_2$ structure function for $Q^2 {\raisebox{-0.7ex}{$\stackrel {\textstyle<}{\sim}$}} 7$\,GeV$^2$. 

The present results provide a consistency check of perturbative QCD although 
they are limited by the available statistics and systematic uncertainties. 
About 160,000 tagged charm events, compared to the present data sample
of a few hundred events, are expected for an integrated 
luminosity of 500\,pb$^{-1}$ and improved tracking (including silicon micro 
vertex detectors), which will allow a detailed study of the dynamics of charm 
production~\cite{daum}. 
The large integrated luminosity anticipated will also give access to the 
bottom structure function $F_2^{b\bar b}$. 

\subsection{Events with Isolated Lepton and Missing $p_t$}

Both collaborations searched for events with isolated leptons and missing 
transverse momentum. 

The H1 analysis~\cite{H1_muon_eps} is based on the 1994 to 1997 summer data 
sample, which corresponds to an integrated luminosity 
of 24.7\,pb$^{-1}$. 
The basic requirement for the event selection was a missing transverse 
momentum of $p_T^{cal} > 25$\,GeV measured in the calorimeter. 
Beam gas background, cosmic and halo muon events were rejected by requiring 
a reconstructed vertex in the interaction region and by using a 
set of topological and timing filters. 
The events were required to have a track with transverse momentum higher 
than 10\,GeV and polar angle greater than $10^\circ$. 
After these cuts 336 events were left. 
Five of these events have a well isolated track; 
3 events contain an isolated muon, 
1 event contains an isolated electron and 1 event contains both an 
isolated positron and an isolated muon. 
%

The main Standard Model process expected to yield sizeable rates of
events with missing transverse momentum and a high $p_T$ isolated 
lepton is the  production and subsequent leptonic decay of a $W$ boson. 
The expected number of events is $1.34\pm0.20$ with isolated electron 
and $0.41\pm0.07$ with an isolated muon. 
The event rate coming from other sources, e.g.~production of heavy quark 
pair (charm or bottom) with subsequent semileptonic decay of one of the 
quarks and production of $\tau$ pairs, was estimated to be less than 
0.05 events. 

\begin{figure}[h]
\centerline{\epsfig{file=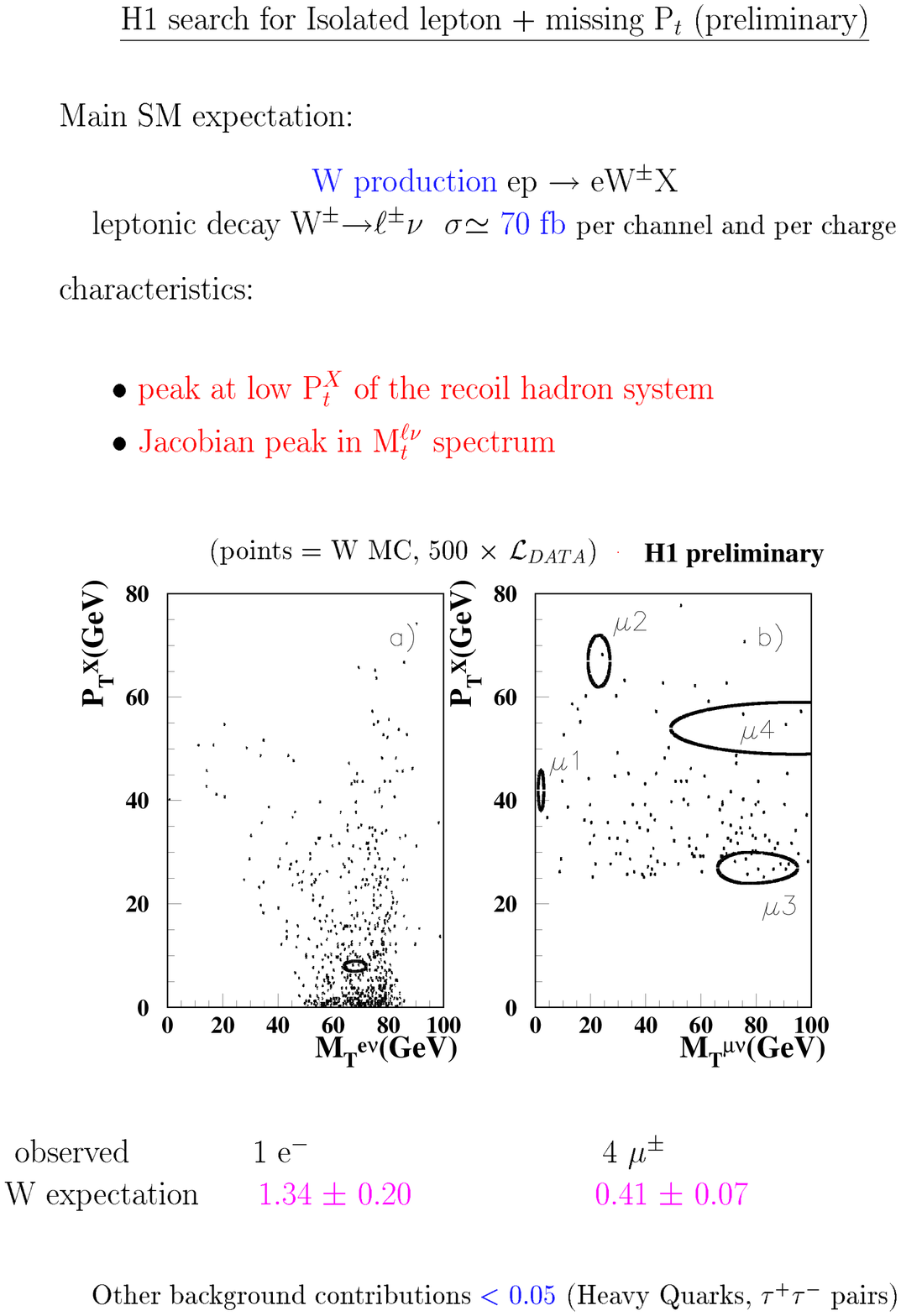,%
             bbllx=80,bblly=180,bburx=480,bbury=450,width=12cm,clip=}}
\caption{\sf $p^X_T$ and $M^{l\nu}_T$ distributions of the observed H1 events:
         a) electron channel, b) muon channel. The elliptic contours correspond
         to the one-sigma error on the measured parameters. 
         The points show the Standard Model expectation for $W$ production 
         with 500 times higher integrated luminosity than the data. }
\label{H1_ptmt}
\end{figure}

Fig.~\ref{H1_ptmt} compares the hadronic transverse momentum $p^X_T$ and 
transverse mass $M^{l\nu}_T$ with the expected distribution of $W$ events. 
The Monte Carlo corresponds to a 500 times higher luminosity than the data. 


A similar analysis was carried out by the ZEUS collaboration. 
The main selection criteria were: energy of the electron candidate $>$ 15\,GeV 
and missing $p_T > 19$\,GeV. For muon candidates a minimum ionizing 
particle with track momentum greater than 5\,GeV and missing $p_T > 18$\,GeV 
was required. 
In addition, there were cuts on the polar angle and quality of the track. 

Using the full 1994 to 1997 data sample of 46.6\,pb$^{-1}$, 
4 events with an isolated electron 
and no events with isolated muon were found. 
The number of expected events in the electron channel are: 
$2.2\pm0.02$ from $W$ production, $0.32\pm0.14$ from neutral current 
deep inelastic scattering, 
$0.65\pm0.17$ from charged current deep inelastic scattering, 
$0.27\pm0.27$ from elastic and inelastic photoproduction ($\gamma \gamma 
              \rightarrow e^+ e^-$) and
$<0.06$ from elastic and inelastic photoproduction ($\gamma \gamma 
              \rightarrow \tau^+ \tau^-$). 
In the muon channel $0.46\pm0.02$ events are expected from $W$ production, 
$<0.06$ from NC DIS, $0.37\pm0.13$ from CC DIS, 
$0.41\pm0.18$ from elastic and inelastic photoproduction ($\gamma \gamma 
              \rightarrow \mu^+ \mu^-$) and
$0.06\pm0.06$ from elastic and inelastic photoproduction ($\gamma \gamma 
              \rightarrow \tau^+ \tau^-$). 
For both channels the observed rate of events is consistent with $W$ 
production. 
In the $\mu$-channel a limit on the $W$ cross section of 
$\sigma(W)(p_T^{had} > 30$\,GeV)$ < 1.2$\,pb$^{-1}$ at 95\,\% CL was obtained. 

%

\subsection{Determination of the $W$ Boson Mass}

The cross section for charged current electron proton scattering can 
be written as a function of the $W$ boson mass $(m_W)$:
\beq
\frac{d^2\sigma^{e^- p} }{dx\ dQ^2}=
\frac{G_F^2}{2\pi x}
{\left({{m_W^2}\over{m_W^2+Q^2}}\right)}^2
\sum_{i=1}^3
\left[
x u_i(x,Q^2)+(1-y)^2x {\overline d}_i(x,Q^2)
\right], 
\eeq
with
$ G_F = \frac{\pi \alpha}{\sqrt{2} m_W^2(1- m_W^2/M_Z^2)} 
\frac{1}{1 - \Delta r(m_t)}, $ 
where the mass of the top-quark $(m_t)$ enters via radiative corrections. 
Fitting the cross section to the measured CC cross section with $m_W$ 
as the only  free parameter, 
the following results were obtained for the $W$ mass: 
H1: $84^{+9+5}_{-6-4}$\,GeV  (including 1994 data, 
3.1 pb$^{-1}$)~\cite{H1CC}, 
ZEUS: $79^{+8+6}_{-7-3}$\,GeV  (including 1994 data, 
3.7 pb$^{-1}$)~\cite{ZEUSCC} and
ZEUS: $78.6^{+2.5+3.3}_{-2.4-3.0}$\,GeV  (including 1997 data, 46.6 pb$^{-1}$).
These results, although $W$ production in the t-channel, 
are consistent with the LEP and Tevatron measurements, 
but currently much less precise. 

A detailed study of the expected sensitivity for $m_W$ with high luminosity 
and polarized electron/positron beam was done by 
R.\,Beyer et al.~\cite{EWstudy}. 
The CC and NC cross sections will be measured as a function of $Q^2$ and 
compared to the Standard Model prediction with $m_W$ and $m_t$ 
as free parameters. 
There is only a weak logarithmic dependence on the Higgs boson mass, 
since it enters via loops.
The CC cross section is more sensitive to $m_W$, as its normalization 
directly depends on $m_W$, whereas in the case of NC it is the shape 
of the cross section as a function of $Q^2$. 
The sensitivity of the measurement is significantly improved with polarized 
electron or positron beams. 
A beam polarization of 70\% is roughly equivalent to a factor of 4 increase in 
integrated luminosity for unpolarized beams. \hfill

\begin{wrapfigure}{r}{8.5cm}
\centerline{\epsfig{file=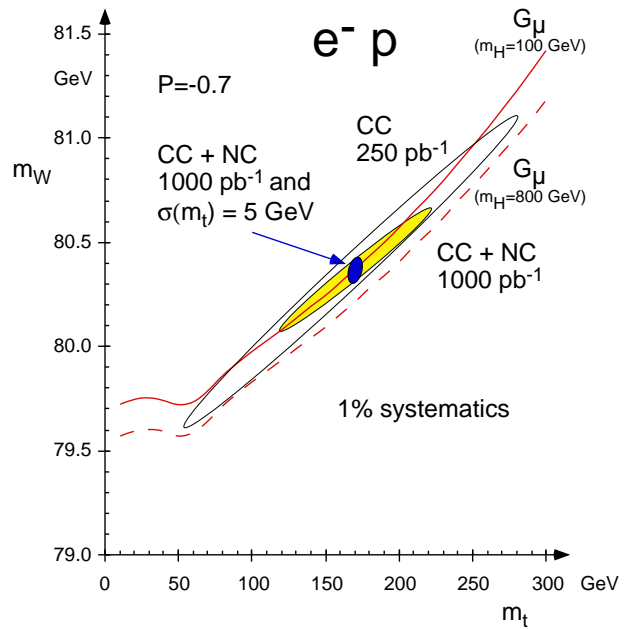,height=8.5cm}}
\caption{\sf Expected measurement of $m_W$, 1 $\sigma$ contours in the 
         $(m_W, m_t)$ plane. }
\label{Wmass}
\end{wrapfigure}

The expected error on $m_W$ was studied by varying the relative systematic
uncertainty from 0 to 5\%. 
The goal is a relative systematic error of 1\%. 
A 1\% determination of the luminosity  has already been achieved. 
A 2\% measurement error of the polarization is conceivable and adequate, since 
it enters through a factor $( 1 + |P|)$ in the cross section. 
The largest experimental uncertainty will be the absolute energy scale 
of the calorimeter, which is presently 1 to 2\% for electrons and 
3\% for hadrons. 
High luminosity data samples will allow to impose constraints on the 
reconstruction of the final state. 
Similar to the present CC analysis, the CC data will be cross checked with 
NC events.

Fig.~\ref{Wmass} shows the expected measurement of $m_W$. 
The 1 $\sigma$ contour is represented by the shaded ellipse. 
The resulting precisions will be $\delta m_W = \pm 300$\,MeV  and 
$\delta m_t = \pm 50$\,GeV. 
Assuming a top-quark mass measurement of CDF and D0 with 5\,GeV uncertainty, 
the expected error is $\delta m_W = \pm 55$\,MeV for 
each experiment assuming a 1\% systematic error. 
The accuracy deteriorates to  $\pm 81$\,MeV for a 2\% systematic uncertainty. 
The expected precision is similar to the expected combined LEP2 measurement 
error of $<$ 50\,MeV  for 500\,pb$^{-1}$~\cite{Thomson}.

\subsection{Search for anomalous $W W \gamma$ Couplings}

In general, the cross section for $e p \rightarrow e W X$ can be written 
as sum of the contributions from the Standard Model, the anomalous 
coupling and an interference term: 
$\sigma_{tot} = \sigma_{SM} + a \sigma_{int} + a^2 \sigma_{an}$, 
where $a = \Delta \kappa, \lambda$ are the anomalous couplings of the 
three-boson vertex $W W \gamma$. 
At HERA a  measurement of the anomalous couplings was done by  ZEUS 
using an integrated luminosity of about 10\,pb$^{-1}$, resulting in 
limits of $\Delta \kappa < 7 (\lambda = 0)$ and 
$ |\lambda| < 11.7 (\Delta \kappa = 0)$ at 95\% CL~\cite{waters}. 

\begin{figure}[h]
\centerline{\epsfig{file=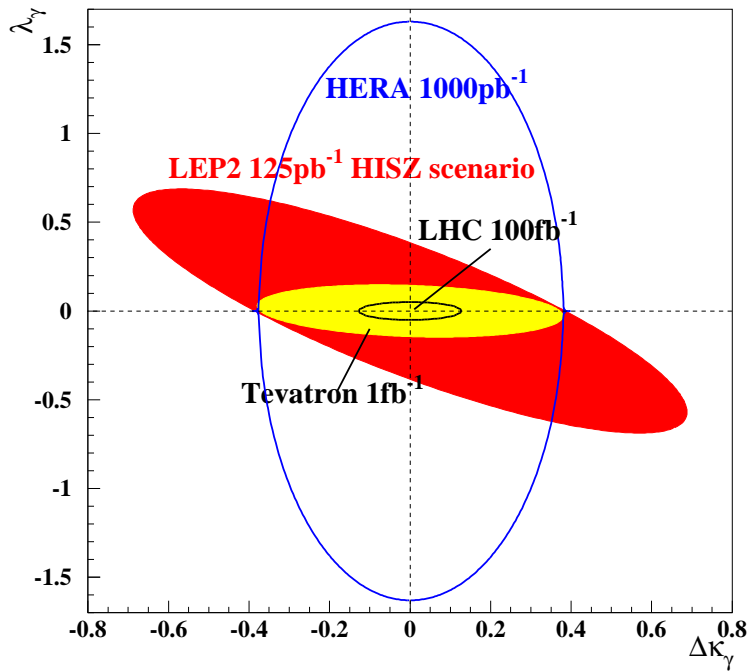,width=10cm}}
\caption{\sf Expected 95\% CL limits for $ W W \gamma$ couplings determined 
from single $W$ production at HERA, $W W$ production at LEP2 and 
$W \gamma$ production at the Tevatron and LHC.}
\label{kappa}
\end{figure}

Fig.~\ref{kappa}~\cite{Noyes} compares the 95\% CL sensitivity limits that 
can be  achieved at HERA  with an integrated luminosity of 1\,fb$^{-1}$  
to projected limits~\cite{ANlimits} from LEP2, Tevatron and LHC. 
HERA results on $\Delta \kappa $ will be competitive to LEP2 and 
Tevatron measurements. 
%
%
\newpage
\subsection{Search for Selectron and Squark Production}

\begin{wrapfigure}[43]{r}{8cm}
\centerline{\epsfig{file=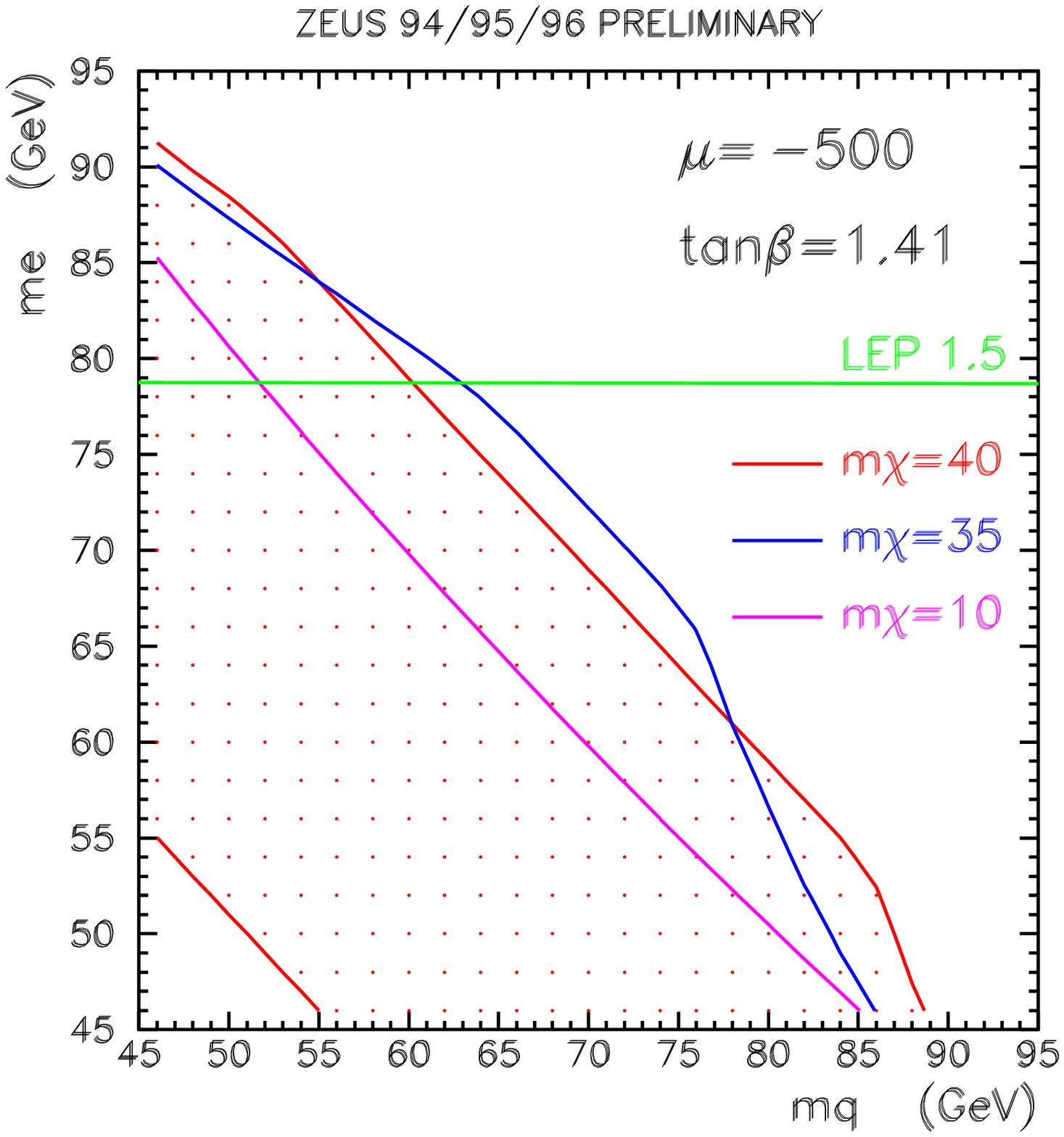,width=8cm}}
\caption{\sf Excluded region at 95\% CL 
 in the plane defined by $m_{\tilde e}$ and $m_{\tilde q}$ for fixed values 
of $m_{\chi_1^0} = 10, 35,$ and 40\,GeV, $\mu = -500$\,GeV and 
$\tan \beta = 1.41$. LEP2 limits ($m_{\chi_1^0} = 40$\,GeV, $\mu = -500$\,GeV,
 $\tan \beta = 1.5$) are shown as a dotted line. }
\label{SUSY1}
%
\centerline{\epsfig{file=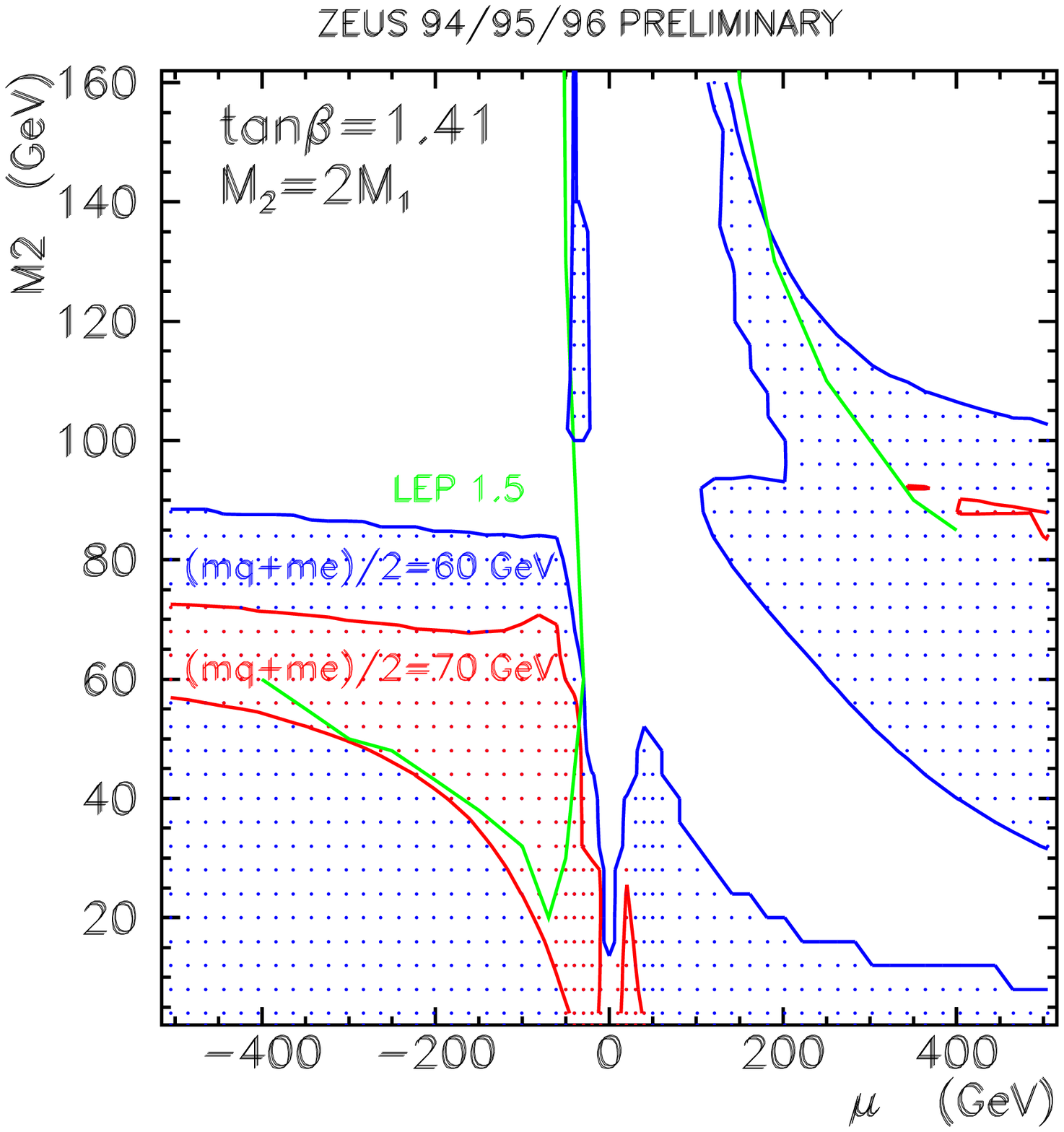,width=8cm}}
\caption{\sf Excluded region at 95\% CL 
 in the plane defined by $M_2$ and $\mu$ for 
$(m_{\tilde e} + m_{\tilde q})/2 = 60$\,GeV (full line) and 70\,GeV 
(dashed line) for $\tan \beta = 1.41$. 
The area below the dotted line is excluded by LEP2. }
\label{SUSY2}
\end{wrapfigure}

A search for selectron and squark production at HERA 
$e p \rightarrow \tilde e_{L,R} \tilde q_{L,R}^f X$ 
was first performed by H1 using the 1994-1995 data sample 
$(\cal{L}$ = 6.4\,pb$^{-1}$)~\cite{H1SUSY}.
ZEUS extended the sensitivity by using the 1994-1996 data sample, 
corresponding to $\cal{L}$ = $(20.0 \pm 0.2)$\,pb$^{-1}$~\cite{ZEUSSUSY}. 
Selectron, squark event candidates  were selected  requiring a well 
identified electron, 
the hadronic system and the electron  not back-to-back in azimuth, 
and missing momentum in the final state. 
Two events survived the selection, in agreement with the expected 
Standard Model backgrounds of $2.74 \pm 0.47$ events. 

Limits on supersymmetric models were calculated assuming that the 
right and left-handed selectrons have the same mass  $m_{\tilde e}$ and 
all quarks (except stop) have the same mass $m_{\tilde q}$ and there 
is no mixing between the L-R scalar fermions. 
Fig.~\ref{SUSY1} shows the region excluded by ZEUS in the plane defined by 
$m_{\tilde e}$ and $m_{\tilde q}$ for fixed values of $m_{\chi_1^0}$. 
For a selectron of $m_{\tilde e} = 80$\,GeV (the LEP2 limit~\cite{SUSYLEP}) 
and a neutralino  of $m_{\chi_1^0} = 40$\,GeV, squarks up to 
$m_{\tilde q}\simeq 60$\,GeV are excluded. 
These limits on squark masses are complementary to those from CDF and 
D0~\cite{SUSYTev}, because the HERA results make no assumptions on the 
gluino mass. 
Fig.~\ref{SUSY2} compares the excluded regions in the $M_2$ versus 
$\mu$ plane,
 for fixed selectron and squark mass, with limits from chargino and 
neutralino searches at LEP~\cite{SUSYLEP2}.


%
High luminosity data will increase the sensitivity to higher masses, 
e.g. limits of $(m_{\tilde e} + m_{\tilde q})/2 = 97$\,GeV for 
$M_2 = 130$\,GeV~\cite{Schleper}. 

\section{Conclusions} 
Electron proton collisions at HERA provide a unique environment for the 
study of QCD and electroweak aspects of the Standard Model. 
So far these measurements were mainly done at low and medium $x$ and $Q^2$. 
High $Q^2$ physics has just begun.  
The obtained results on searches beyond the Standard Model are  
competitive with other measurements. 
In 1998/1999 a $e^- p$ data sample of about 50\,pb$^{-1}$ is expected. 
The present physics program  will continue which the full detector 
upgrades. 
In addition, comparisons of $e^- p$ and $ e^+ p$ cross sections will be 
performed. 
After the upgrade of the machine in 2000, a factor of $\approx 5$ increase 
in luminosity and a total integrated luminosity from 5 years running 
of $\approx $1\,fb$^{-1}$ per experiment is expected, 
which will  provide a very rich physics program.

\end{document}